\documentclass{llncs}

\usepackage[latin1]{inputenc}
\usepackage{graphicx}
\usepackage{algorithm, algorithmic}
\def\thealg{\textit{\textbf{PROSA\/\/ }}}

\mainmatter

\title{Exploiting social networks dynamics for P2P resource organisation}
\titlerunning{Exploiting social networks dynamics for P2P resource organisation}
\author{Vincenza Carchiolo\inst{1} \and Michele Malgeri\inst{1} \and
Giuseppe Mangioni\inst{1} \and Vincenzo Nicosia\inst{1}}
\tocauthor{Vincenza Carchiolo (Universit\`a di Catania),
Michele Malgeri (Universit\`a di Catania),
Giuseppe Mangioni (Universit\`a di Catania),
Vincenzo Nicosia (Universit\`a di Catania)}
\institute{Dipartimento di Ingegneria Informatica e delle 
Telecomunicazioni \\
Facolt\`a di Ingegneria -- Universit\`a di Catania\\
Viale A. Doria 6 -- 95100 Catania (ITALY)\\
\email{\{car,malgeri,gmangioni,vnicosia\}@diit.unict.it}
}

\begin{document}
\maketitle

\abstract{
In this paper we present a formal description of \thealg, a P2P
resource management system heavily inspired by social networks. 
Social networks have been deeply studied in the last two decades in
order to understand how communities of people arise and grow. It is a
widely known result that networks of social relationships usually
evolves to small--worlds, i.e. networks where nodes are strongly
connected to neighbours and separated from  all other nodes by a small
amount of hops. This work shows that algorithms implemented into
\thealg allow to obtain an efficient small--world P2P network.}

\section{Introduction}
\label{s:intro}

A Peer-to-Peer system consists of computing elements that are
connected by a network, addressable in a unique way, and sharing a
common communication protocol. All computing elements, equivalently
called nodes or peers, have the same functionalities and role. In P2P
networks there is no difference between "client" and "server" hosts: a
peer acts as a "client" if it requests a resource from the network,
while it acts as a "server" if it is requested a resource it is
sharing. From this point of view, P2P networks differ a lot from World
Wide Web, TCP/IP networks and, in general, from client-server
networks.

Studies on P2P networks are focused on two different topics: physical
P2P networks (i.e.,P2P networks opposed to hierarchic and centralised
TCP/IP networks) and overlay networks (i.e. networks of logical links
between hosts over an existing physical network of any type).  Our
interest is mainly focused on overlay P2P systems: they are probably
going to become the most used kind of application--level protocols
for resource sharing and organisation.

In this paper we present a novel P2P overlay network, named \thealg,
heavily inspired by social networks. Social networks are sets
of people or groups interconnected by means of acquaintance,
interaction, friendship or collaboration links. Many kinds of natural
social networks have been deeply studied in the last thirty years
\cite{nexus-buchanan}, and many interesting characteristics of such
networks have been discovered. In a real social network relationships
among people are of the most importance to guarantee efficient
collaboration, resources discovery and fast retrieval of remote
people. Nevertheless, not all relationships in a social network are of
the same importance: usually links to parents and relatives are
stronger than links to friends, which are in turn stronger than links
to colleagues and class mets. On the other hand, it is also
interesting to note that usually links in a social group evolve in
different ways. A large amount of relationships are (and remain) bare
``acquaintances''; some of them evolve around time into
``friendships'', while ``relativeness'' is typical of very strong
links to really trusted people.

This suggests that a P2P network based on a social model should take
into account that different kind of links among peers can exist, and
that links can evolve from simple acquaintances to friendship.

Results of studies performed by Watts, Strogatz, Newman, Barabasi et
al. in the last decades\cite{watts98strogatz} \cite{scott-handbook}
\cite{newman-2001-98} \cite{barabasi-food-2002} 
\cite{albert-2002-74}
reveal that networks of movie characters, scientific collaborations,
food chains, proteins dependence, computers, web pages and many other
natural networks usually exhibit emerging properties, such that of
being small--worlds.  A small--world is a network where distance
among nodes grows as a logarithmic function of the network size and
similar nodes are strongly connected in clusters.

\thealg tries to build a P2P network based on social relationships, in
the hope that such network could naturally evolve to a small--world. 

In section \ref{s:prosa}, we describe \thealg and algorithms involved
in linking peers and routing queries for resources; in section
\ref{sect:topo}, we report some results about topological properties of
\thealg network, obtained by simulation; in section
\ref{sect:future}, we summarise obtained results and plan future work.

\section{\thealg: A brief introduction}
\label{s:prosa}
As stated above, \thealg{} is a P2P network based on social
relationships.  More formally, we can model \thealg as a directed
graph: \begin{equation} \label{equ:prosa} \thealg = (\mathcal{P},
\mathcal{L}, P_k, Label) \end{equation}

$\mathcal{P}$ denotes the set of peers (i.e. vertices), $\mathcal{L}$
is the set of links $l = (s, t)$ (i.e. edges), where $t$ is a neighbour
of $s$.  For link $l = (s, t)$, $s$ is the source peer and $t$ is the
target peer. All links are directed.

In P2P networks the knowledge of a peer is represented by the
resources it shares with other peers.  In \thealg the mapping $P_k:
\mathcal{P} \rightarrow \mathcal{C}$, associates peers with
resources. For a given peer $s \in \mathcal{P}$, $P_k(s)$ is a compact
description of the peer knowledge ($PK$ - {\em Peer Knowledge}).

Relationships among people are usually based on similarities in
interests, culture, hobbies, knowledge and so on. Usually these kind
of links evolve from simple ``acquaintance--links'' to what we called
``semantic--links''.  To implement this behaviour three types of links
have been introduced: {\em Acquaintance--Link} ($AL$), {\em Temporary
Semantic--Link} ($TSL$) and {\em Full Semantic--Link} ($FSL$).  TSLs
represent relationships based on a partial knowledge of a peer. They
are usually stronger than $AL$s and weaker than $FSL$s.

In \thealg, if a given link is a simple $AL$, it means that the source
peer does not know anything about the target peer.  If the link is a
$FSL$, the source peer is aware of the kind of knowledge owned by the
target peer (i.e. it knows the $P_k (t)<$, where $ \;t \in
\mathcal{P}$ is the target peer). Finally, if the link is a $TSL$, the
peer does not know the full $P_k(t)$ of the linked peer; it instead
has a {\em Temporary Peer Knowledge} ($TP_k$) which is built based on
previously received queries from the source peer.  Different meanings
of links are modelled by means of a labelling function $Label$: for a
given link $l = (s,t) \in L$, $Label(l)$ is a vector of two elements
$[e, w]$: the former is the link label and the latter is a weight used
to model what the source peer knows of the target peer; this is
computed as follow:
\begin{itemize}
	\item if $e = AL \Rightarrow w = \emptyset$ 
	\item if $e = TSL \Rightarrow w = TP_k$
	\item if $e = FSL \Rightarrow w = P_k(t)$ 
\end{itemize}

In the next two sections, we give a brief description of how
\thealg works.  A detailed description of \thealg{} can be found in
\cite{prosa@cops06}.


\subsection{Peer Joining to \thealg{}}
The case of a node that wants to join an existing network is similar
to the birth of a child. At the beginning of his life a child
``knows'' just a couple of people (his parents). A new peer which
wants to join, just looks for $n$ peers at random and establishes
$AL$s to them.  These links are $AL$s because a new peer doesn't know
anything about its neighbours until he doesn't ask them for
resources. This behaviour is quite easy to understand: when a baby
comes to life he doesn't know anything about his parents.  The
\thealg{} peer joining procedure is represented in algorithm
\ref{join}.

\begin{algorithm}
\caption{\small {\bf JOIN:} Peer $s$ joining to $\thealg (\mathcal{P},\mathcal{L}, P_k, Label)$}
\label{join}
\begin{algorithmic}[1]

	\REQUIRE $\thealg (\mathcal{P}, \mathcal{L}, P_k, Label), Peer\:s$
	\STATE $\mathcal{RP} \leftarrow rnd (P, n)$ \COMMENT {Randomly selects $n$ peers of \thealg}
	\STATE $\mathcal{P} \leftarrow \mathcal{P} \cup s$ \COMMENT {Adds $s$ to set of peers}
	\STATE $\mathcal{L} \leftarrow \mathcal{L} \cup \{(s, t), \forall t \in \mathcal{RP}\}$ 
		   \COMMENT {Links $s$ with the randomly selected peers}
	\STATE $\forall t \in \mathcal{RP} \Rightarrow Label(p, q) \leftarrow [AL, \emptyset]$
			\COMMENT {Sets the previously added links as $AL$}
\end{algorithmic}
\end{algorithm}

\subsection{\thealg{} dynamics}
In order to show how \thealg{} works, we need to define the structure
of a query message.  Each query message is a quadruple:
\begin{equation}
	Q_M = (qid, q, s, n_r)
\end{equation}

where $qid$ is a unique query identifier to ensure that a peer does
not respond to a query more then once; $q$ is the query, expressed
according to the used knowledge model\footnote{If knowledge is
modelled by Vector Space Model, for example, $q$ is a state vector of
stemmed terms. If knowledge is modelled by onthologies, $q$ is an
ontological query, and so on}; $s \in P$ is the source peer and $n_r$
is the number of required results.
\thealg dynamic behaviour is modelled by Algorithm \ref{search} and
is strictly related to queries.  When a user ($U$) of \thealg asks for
a resource on a peer $s$, the inquired peer $s$ builds up a query $q$
and specify a certain number of results he wants to obtain $n_r$. This
is equivalent to call $ExecQuery (\thealg, s,
\emptyset, n_r)$.

\begin{algorithm}
\caption{\footnotesize{\bf ExecQuery:} query $q$ originating from peer $s$ executed on peer $cur$}
\label{search}
\begin{algorithmic}[1]
\small
	\REQUIRE $\thealg(\mathcal{P}, \mathcal{L}, P_k, Label)$
	\REQUIRE $cur, prev  \in \mathcal{P}, \;qm \in QueryMessage$
	\STATE $Result \leftarrow \emptyset $
	\IF {$prev \neq \emptyset$}
		\STATE $UpdateLink (\thealg, cur, prev, q)$
	\ENDIF
	\STATE $(Result,numRes) \leftarrow ResourcesRelevance (\thealg{}, q, cur, n_r)$
	\IF {$numRes = 0$}
		\STATE $f \rightarrow SelectForwarder(\thealg, cur, q)$
		\IF {$f \neq \emptyset$}
			\STATE $ExecQuery (\thealg, f, cur, qm)$
		\ENDIF
	\ELSE
		\STATE $SendMessage (s, cur, Result)$
		\STATE $\mathcal{L} \leftarrow \mathcal{L} \cup (s, cur)$
		\STATE $Label (s, cur) \leftarrow [FSL, P_k(cur)]$
		\IF {$numRes < n_r$}
			\STATE \COMMENT {-- Semantic Flooding --}
			\FORALL {$t \in Neighborhood(cur)$}
				\STATE $rel \rightarrow PeerRelevance(P_{k}(t), q)$
				\IF {$rel > Threshold$}
					\STATE $qm \leftarrow (qid, q, s, n_r - numRes)$
					\STATE $ExecQuery (\thealg, t, cur, qm)$
				\ENDIF
			\ENDFOR
		\ENDIF
	\ENDIF
\end{algorithmic}
\end{algorithm}

The first time $ExecQuery$ is called, $prev$ is equal to $\emptyset$
and this avoids the execution of instruction number $3$.  Following
calls of $ExecQuery$, i.e. when a peer receives a query forwarded by
another peer, use function $UpdateLink$, which updates the link
between current peer $cur$ and the forwarding peer $prev$, if
necessary. If the requesting peer is an unknown peer, a new $TSL$ link
to that peer is added having as weight a Temporary Peer
Knowledge($TP_k$) based on the received query message. Note that a
$TP_k$ can be considered as a ``good hint'' for the current peer, in
order to gain links to other remote peers. It is really probable that
the query would be finally answered by some other peer and that the
requesting peer will download all resources that matched it. It would
be useful to record a link to that peer, just in case that kind of
resources would be requested in the future by other peers.  If the
requesting peer is a $TSL$ for the peer that receives the query, the
corresponding TPV (Temporary Peer Vector) in the list is updated. If
the requesting peer is a $FSL$, no updates are performed.

The relevance of a query with respect to the resources hosted by the
user's peer is evaluated calling function $ResourcesRelevance$.  Two
possible cases can hold: 
\begin{itemize}
\item
 If none of the hosted resources has a
sufficient relevance, the query has to be forwarded to another peer
$f$, called ``forwarder''.  This peer is selected among $s$ neighbours
by the function $SelectForwarder$, using the following procedure:
	
	\begin{itemize}
	\item[-]
	Peer $s$ computes the relevance between query $q$ and the weight
	of each links connecting itself to his neighbourhood.
	\item[-]
	It selects the link with the highest relevance, if any, and
	forward the query message to it.
	\item[-]
	If the peer has neither $FSL$s nor $TSL$s, i.e. it has just $AL$s,
	the query message is forwarded to one link at random.
	\end{itemize}
This procedure is described in Algorithm \ref{search}, where the
subsequent forwards are performed by means of recursive calls to
$ExecQuery$.
	
\item
If the peer hosts resources with sufficient relevance with
respect to $q$, two sub-cases are possible:

	\begin{itemize}
	\item[-] The peer has sufficient relevant documents to full-fill the
	request. In this case a result message is sent to the requesting
	peer and the query is no more forwarded.
	\item[-] The peer has a certain number of relevant documents, but
	they are not enough to full-fill the request (i.e. they are $<
	n_r$).  In this case a response message is sent to the requester
	peer, specifying the number of matching documents. The message query
	is forwarded to all the links in the neighbourhood whose relevance
	with the query is higher than a given threshold (semantic
	flooding). The number of matched resources is subtracted from the
	number of total requested documents before each forward step.
  \end{itemize}
\end{itemize}

	When the requesting peer receives a response message it presents the
	results to the user. If the user decides to download a certain
	resource from another peer, the requesting peer contacts the peer
	owning that resource asking for download. If download is
	accepted, the resource is sent to the requesting peer.

\section{Topological properties}
\label{sect:topo}





Algorithms described in section \ref{s:prosa} are inspired by the way
social relationships among people evolve, in the hope that a network
based on those simple rules could naturally become a small--world.
That of being a small--world is one of the most desirable properties
of a P2P network, since resource retrieval in small--worlds is really
efficient.  This is mainly due to the fact that small--world networks
have a short Average Path Length (APL) and a high Clustering
Coefficient (CC).  APL is defined as the average number of hops
required to reach any other node in the network: if APL is small, all
nodes of the network can be easily reached in a few steps starting
from whichever other node.

CC can be defined in several ways, depending on the kind of
``clustering'' you are referring to.  We used the definition given in
\cite{watts98strogatz}, where the clustering coefficient of a node is
defined as:

\begin{equation}
	CC_{n} = \frac{E_{n,real}}{E_{n,tot}}
\label{eq:ccnode}
\end{equation}

where $n$'s neighbours are all the peers to which $n$ as linked to,
$E_{n,real}$ is the number of edges between n's neighbours and
$E_{n,tot}$ is the maximum number of possible edges between n's
neighbours. Note that if $k$ is in the neighbourhood of $n$, the
vice-versa is not guaranteed, due to the fact that links are directed.
The clustering coefficient of the whole network is defined as:

\begin{equation}
	CC = \frac{1}{|V|}\sum_{n \in V}{CC_{n}}
\end{equation}

i.e. the average clustering coefficient over all nodes.

The CC is an estimate of how strongly nodes are connected to each
other and to their neighbourhood. In particular, the definition given
in Equation \ref{eq:ccnode} measures the percentage of links among a
node neighbours with respect to the total possible number of links
among them.

In the following two subsections we show that \thealg has both a small
APL and a considerable high CC.

\subsection{Average path length}
\label{subsect:APL}

Since we are focusing on topological properties of a \thealg network
to show that it is a small--world (i.e. that queries in \thealg are
answered in a small amount of steps), we estimate the APL as the
average length of the path traversed by a query.  It is interesting to
compare the APL of \thealg with the APL of a correspondent random
graph, since random graphs usually have a really small average path
length.

Given a graph $G(V, E)$ with $|V|$ vertices (nodes) and $|E|$ edges
(links) among nodes, the correspondent random graph is a graph
$G_{rnd}$ which has the same number of vertices (nodes) and the same
number of edges (links) of $G$, and where each link between two nodes
exist with a probability $p$.

Note that the APL of a random graph can be calculated using equation
(\ref{eq:aplrnd}), as reported in \cite{newman-physrev-2}, where $|V|$
is the number of vertices (nodes) and $|E|$ is the number of edges
(links).

\begin{equation}
  APL = \frac{\log{|V|}}{\log{(|V|/|E|)}}
	\label{eq:aplrnd}
\end{equation}

\begin{figure}[!htbp]
	\centering
	\includegraphics[scale=0.25]{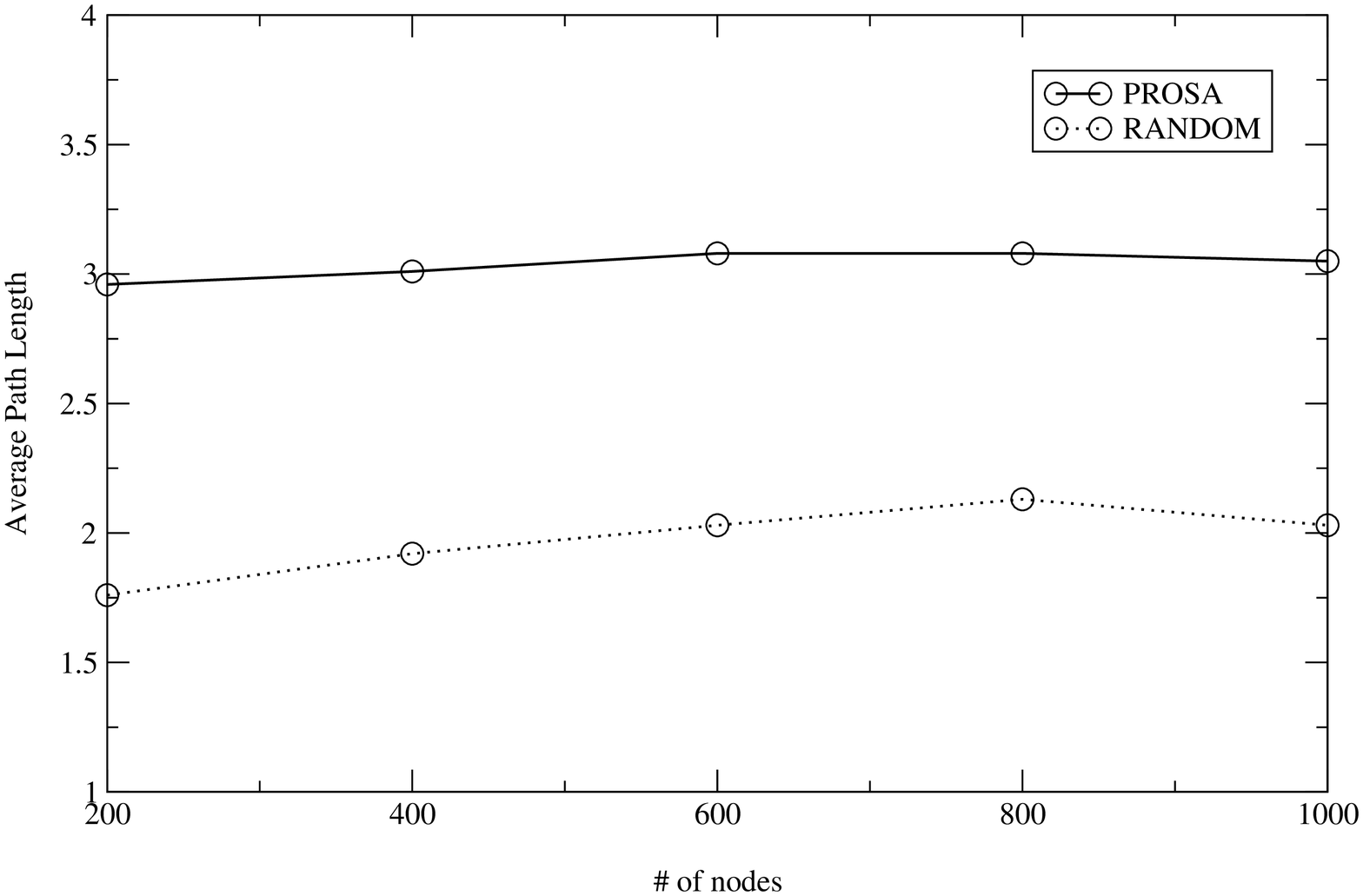}
	\caption{APL for \thealg and random network}
	\label{fig:aplnodes}
\end{figure}

Figure \ref{fig:aplnodes} shows the APL for \thealg and the
correspondent random graph for different number of nodes in the case
of 15 performed queries per node. The APL for \thealg is about 3.0, for
all network sizes, while the APL for the correspondent random graph is
between 1.75 and 2.0: the average distance among peers in \thealg
seems to be independent from the size of the network. This is quite
common in real small--world networks. 

\begin{figure}[!htbp]
	\centering
	\includegraphics[scale=0.25]{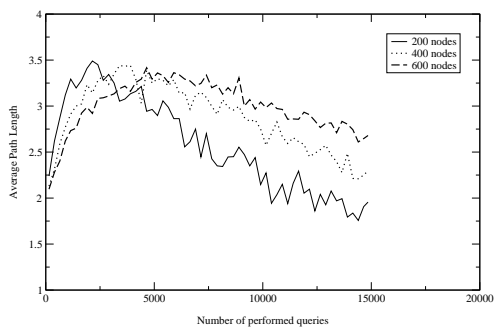}
	\caption{Running averages of APL for \thealg with different network size}
	\label{fig:aplqueries}
\end{figure}

It is also interesting to analyse how APL changes when the total number of
performed queries increases. Results are reported in Figure
\ref{fig:aplqueries}, where the APL is calculated for windows of 300
queries, with an overlap of 50 queries. Note that the APL for \thealg
decreases with the number of performed queries. This behaviour heavily
depends on the facts that new links among nodes arise whenever a new
query is performed (TSLs) or successfully answered (FSLs). The higher
the number of performed queries, the higher the probability that a
link between two nodes does exist.

\subsection{Clustering Coefficient}
\label{subsect:CC}

The clustering (or transitivity) of a network is a measure of how
strongly nodes are connected to their neighbourhood. Since links among
nodes in \thealg are established as a consequence of query forwarding
and answering, we suppose that peers with similar knowledge will be
eventually linked together. This means that usually peers have a
neighbourhood of similar peers, and having strong connections with
neighbours could really speed--up resource retrieval.

\begin{figure}[!htbp]
	\centering
	\includegraphics[scale=0.25]{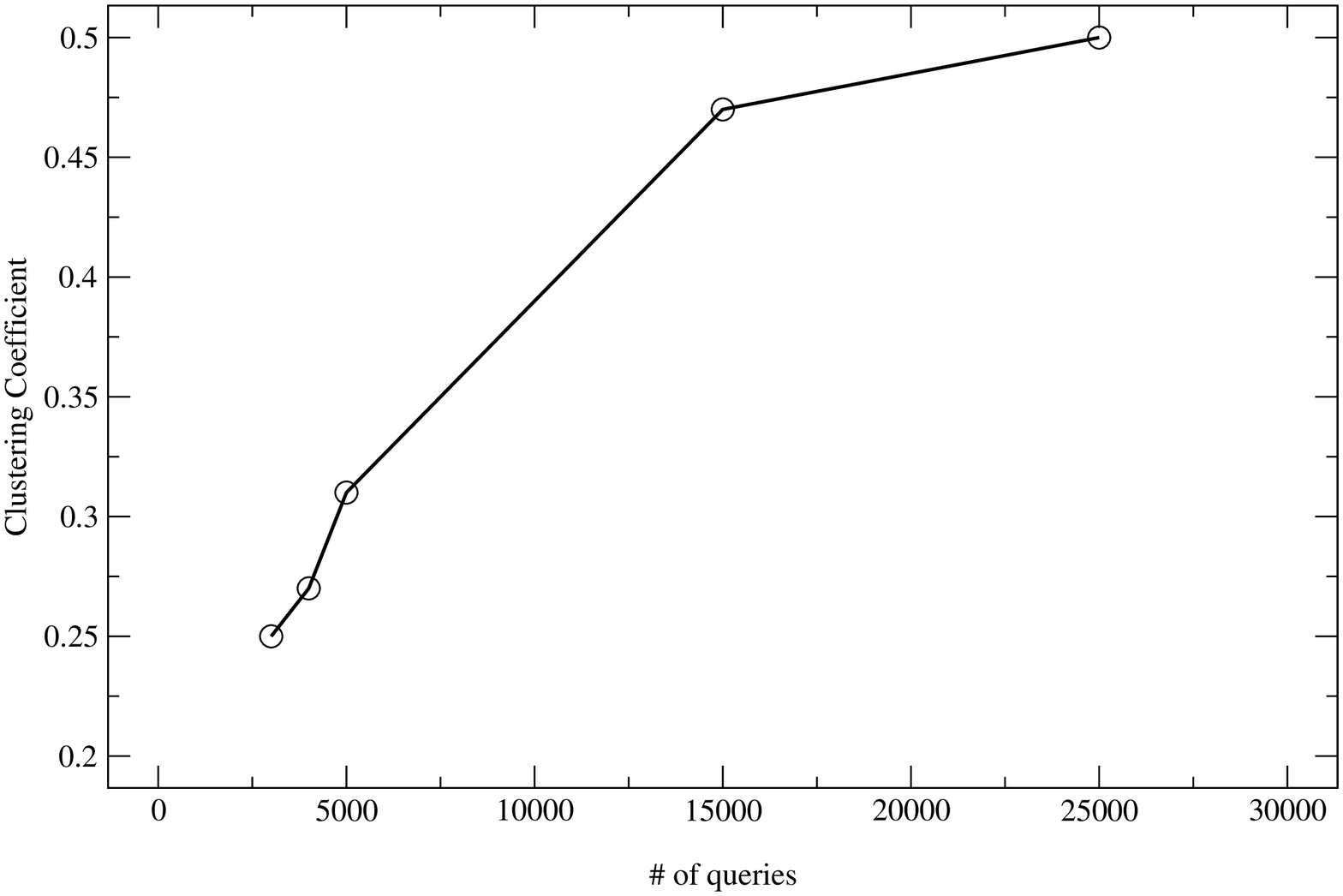}
	\caption{\thealg CC for \thealg}
	\label{fig:ccquery}
\end{figure}

In Figure \ref{fig:ccquery} the CC of \thealg for different number of
performed queries is reported, for a network of 200 nodes. Note that
the clustering coefficient of the network increases when more queries
are performed. This means that nodes in \thealg usually forward
queries to a small number of other peers so that their aggregation
level naturally gets stronger when more queries are issued.

It could be interesting to compare \thealg clustering coefficient with
that of a corresponding random graph.  The clustering coefficient of a
random graph with $|V|$ vertices (nodes) and $|E|$ edges (links) can be
computed using equation \ref{eq:cc}.

\begin{equation}
  CC_{rnd} = \frac{|E|}{|V|\cdot(|V| - 1)}
	\label{eq:cc}
\end{equation}

\begin{figure}[!htbp]
	\centering
	\includegraphics[scale=0.25]{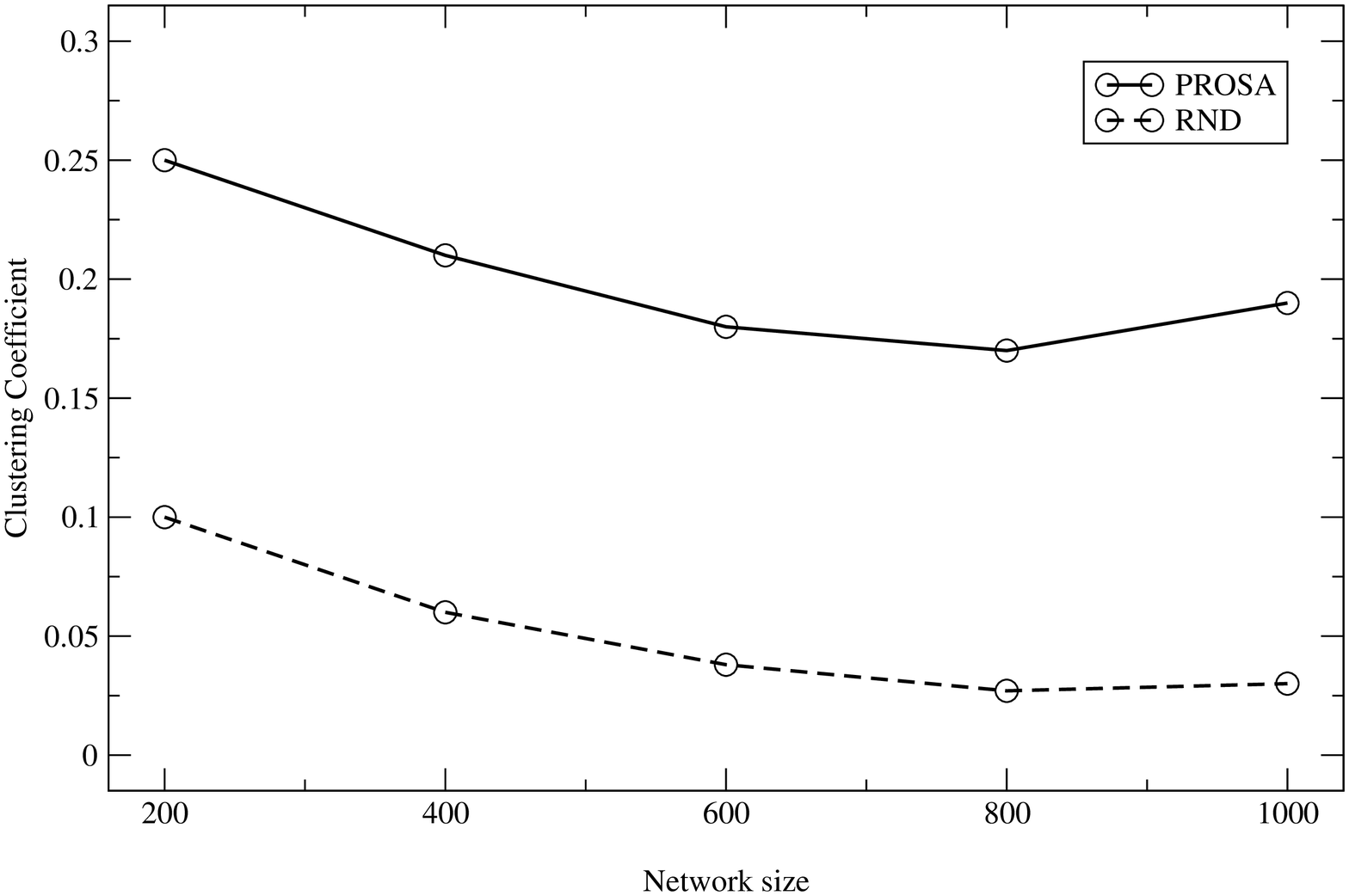}
	\caption{Clustering coefficient for \thealg and the corr. random graph}
	\label{fig:cc}
\end{figure}

Figure \ref{fig:cc} shows the CC for \thealg and a correspondent
random graph for different network sizes, in the case of 15 performed
queries per node. The CC for \thealg is from 2.5 to 6 times higher
that that of a correspondent random graph, in accordance with CC
observed in real small--world networks. This result is quite simple to
explain, since nodes in \thealg are linked principally to similar
peers, i.e. to peers that share the same kind of resources, while
being linked to other peers at random.  Due to the linking strategy
used in \thealg, it is really probable that neighbours of a peer are
also linked together, and this increases the clustering coefficient.

\section{Conclusions and future work}
\label{sect:future}

\thealg is a P2P system mainly inspired by social networks and
behaviours. Topological properties of \thealg suggest that it
naturally evolves to a small--world network, with a very short average
path length and a high clustering coefficient. More results about
query efficiency are reported in \cite{prosa@cops06}. Future work
includes deeply examining the internal structure of \thealg networks
and studying the emergence of communities of similar peers.

\bibliography{article}
\bibliographystyle{plain}
\end{document}